\documentclass[a4paper, 11pt,]{article}
\newcommand{\tabincell}[2]{\begin{tabular}{@{}#1@{}}#2\end{tabular}}  
\usepackage[dvipdfmx]{graphicx}
\usepackage[cmex10]{amsmath} 
\usepackage{mathptmx}
\usepackage{amssymb}
\usepackage{lscape}
\usepackage{times}
\usepackage{here}
\usepackage{multirow}
\usepackage{diagbox}
 \usepackage{booktabs}
\usepackage[normalem]{ulem}
\useunder{\uline}{\ul}{}
\usepackage[usenames]{color}
 \usepackage{caption}

\usepackage{indentfirst}

\title{Saliency difference based objective evaluation method for\\a superimposed screen of the HUD with various background} 

\author{Hailong LIU~$^*$ \and Toshihiro HIRAOKA~$^*$ \and Takatsugu HIRAYAMA~$^*$ \and Dongmin KIM \thanks{Nagoya University, 464-860, JAPAN (e-mail: liuh@murase.m.is.nagoya-u.ac.jp; toshihiro.hiraoka@mirai.nagoya-u.ac.jp; takatsugu.hirayama@nagoya-u.jp; kim.dong.min@b.mbox.nagoya-u.ac.jp)}}
\date{}

\begin{document}
\maketitle
\begin{abstract}               
The head-up display~(HUD) is an emerging device which can project information on a transparent screen. The HUD has been used in airplanes and vehicles, and it is usually placed in front of the operator's view.
In the case of the vehicle, the driver can see not only various information on the HUD but also the backgrounds (driving environment) through the HUD.
However, the projected information on the HUD may interfere with the colors in the background because the HUD is transparent.
For example, a red message on the HUD will be less noticeable when there is an overlap between it and the red brake light from the front vehicle.
As the first step to solve this issue, how to evaluate the mutual interference between the information on the HUD and backgrounds is important.
Therefore, this paper proposes a method to evaluate the mutual interference based on saliency.
It can be evaluated by comparing the HUD part cut from a saliency map of a measured image with the HUD image. 
\end{abstract}

\section{Introduction}
Information of which an in-vehicle information system (IVIS) provides to the driver increases rapidly with the enormous development in information technologies of the vehicle. 
A head-up display~(HUD) is a transparent display to show information from the IVISs such as advanced driver assistance systems~(ADASs)~\cite{thompson2006using} and car-navigation systems~\cite{kazimierski2017integrated}.
The HUD is often placed close to the central vision of the driver, and therefore the driver can recognize both of the information on the HUD and the situation in the background at the same time since the HUD is transparent. 
The reaction time of the driver who received information from the HUD will become faster than the case when the information is shown on a multi-information display on a meter cluster because the distance and the time of the eye movement are reduced. 
Consequently, the HUD has been widely used in the field of aviation, such as civilian aircrafts and fighter jets~\cite{arthur2005flight}.

However, every coin has two sides. The HUD also has obvious disadvantages because of its transparency.
The driver may hardly recognize the projected information on the HUD when the information (e.g. texts, icons, graphs) and the background overlap each other with the same color.
Most of the HUDs used in the airplanes have a single color, e.g., green or orange~\cite{avionics_handbook}, because there are no over-complicated colors in the sky background (white for the clouds, blue for the sky, black for the night sky).
In the case of the vehicle, there are various objects with different colors and different shapes all around, such as buildings, trees, vehicles, pedestrians, even the traffic markers on the road.
The driver must visually recognize various information and objects in the traffic environment. 
The environment on the ground is more complex than that in the sky, and therefore the complex traffic environment on the background might affect the driver's recognition of information on the HUD. 
On the other hand, in the case when the information is too prominent, the projected information on the HUD might deteriorate the visibility of the objects in the background of which the driver needs to recognize.
It will induce a dangerous situation especially in an emergent case such as rush-out of the pedestrian in front of the vehicle. 

Many studies analyzed the effect of the HUD on the driver from different perspectives, such as the eye movement~\cite{weintraub1985head}, the focusing ability~\cite{charissis2007evaluation}, the information visibility~\cite{inuzuka1991visibility}, the response times~\cite{Enomoto201620164196,wolffsohn1998influence}.
The subject evaluation are adopted in most studies. 
They can directly get the participant's feedback, but they require too many costs such as time and manpower to perform the evaluations. 
By contrast, the objective evaluations without subject experiments would save the manpower cost and time cost.
The objective evaluation method generally evaluates the object based on a hypothesis that simulates the participant's feedback. 

The purpose of this paper is to propose a method in order to evaluate the mutual effect between the projected information on the HUD and the background, which is based on a human's visual perception process.
Specifically, this paper proposes a very simple objective evaluation method using a concept of visual saliency.
The visual saliency is an early vision mechanism for finding objects~\cite{itti2000saliency}.
There is a high probability for the driver to notice the projected information on the HUD when its visual saliency is high.
The contributions of this paper are considered as follows:
\begin{itemize}
\item The mutual effect between background and the projected information on the HUD is objectively analyzed.
\item The proposed method would help the HUD designer to evaluate quickly and intuitively whether the HMI is easy to recognize for users under a certain circumstance.
\end{itemize}

\section{Background}
\subsection{Evaluations for HUD}
The HUD was investigated in many studies with respect to its effect on the drivers from different perspectives.
\cite{inuzuka1991visibility} evaluated the effect of the HUD on users from an information cognitive perspective.
They considered that several factors affect the cognition of information on the HUD, such as location, character size, brightness, and color, and investigated the gaze point distribution, the recognition time, and the subjective evaluation when the participants used a HUD. The results showed that the recognition time using HUD is significantly shorter than that using the conventional instrument panel display.
This study clarified the advantages of the HUD, but some studies mentioned that the HUD may be interfered by the recognition of surrounding information when drivers are in a complex background.

\cite{Enomoto201620164196} investigated the reaction time of the driver's braking behavior using the HUD in the case when a brake light of a preceding vehicle is turned on. 
They used a driving simulator to conduct a subject experiment.
The results showed that the brake reaction time of the driver became slower when the projected information on the HUD overlapped with the brake light turned on by the preceding vehicle.
\cite{charissis2007evaluation} used a full-windshield display to analyze the focusing ability of the driver in different focal levels by subject experiments.
Participants encountered difficulties in focusing when the projection distance was shortened.
Most of the participants relied on the HUD to drive in low visibility scenarios such as snow.
In this case, the colors became undiversified and the lines became indistinct in the background.
Meanwhile, the projected information on the HUD becomes more salient.
They also considered that the design of the HUD needs to reduce visual clutter and distraction in high visibility conditions.
\cite{wolffsohn1998influence} analyzed the response time and the detection rate when the HUD image and the outside world scene changed.
They considered that the HUD could reduce the eye movement distance of drivers, but the cognitive demand required by the HUD task may distract the driver from observing the outside world and cause drivers' response times become shorter.
This problem will become more pronounced with age.

\subsection{Saliency map methods}
visual saliency map methods can be roughly divided into two types: a top-down type and a bottom-up type.
If a person has no prior predictions about the appearance of an object or do not actively find it, then it is the bottom-up saliency.
Conversely, if a person takes the initiative to search an object or has prior knowledge of the appearance of the object, then it is the top-down saliency.

For the bottom-up saliency, \cite{itti1998model} proposed a basic method.
It simulates the early vision process of the human when they are attracted attention by a certain object~\cite{itti2000saliency}.
Specifically, the early vision process starts from the light hits the retina and the photoreceptor cells are activated until the ganglion cells are activated.
\cite{itti1998model}'s method considers three elements of visual saliency such as intensity, color, and orientation.
Meanwhile, they mimicked a contrast detection in the receptive field and proposed center-surround differences and an across-scale summation in the method.
On the basis of previous studies, \cite{itti2000saliency} considered the influence of the dynamical change of the image on visual saliency, and added elements of motion in saliency map methods.
\cite{walther2006modeling} proposed a bottom-up saliency method, which is a biologically plausible model of forming and attending to proto-objects in natural scenes. 
This model can be used to serialize the visual processing by the biologically plausible model of object recognition for attending to proto-objects.
\cite{5206596} proposed a bottom-up saliency method that is a frequency tuned approach of computing saliency in images using low-level features of color and luminance.

For the top-down saliency, many models used machine learning approaches to train the parameters of the saliency models.
When training the model, the position of the object in the image is often used as the label, or the eye information of human is directly used as the label.
\cite{kavak2013visual} considered that the saliency estimation could be seen as a supervised learning problem and accordingly applied a machine learning method.
They proposed a top-down saliency model based on multiple bottom-up saliency methods with the fixation maps and used a multiple kernel learning method to train this model.
Deep learning methods were also used for training the top-down saliency model.
\cite{pan2015end} proposed an end-to-end saliency model based on a convolutional network.
They used fixation information as the label like \cite{kavak2013visual}.
This model had a good performance in saliency prediction because the convolutional network had a strong ability to extract local features of images.

For all of the above methods, it is difficult to evaluate the mutual interference of saliency between the HUD and background, if the saliency is computed from only an image including both of them. 
Therefore, this paper proposes a method to deal with the mutual interference, which can be applied to any the top-down and the bottom-up saliency methods. 

\section{Proposed method}
\begin{figure}[tb]
\centering

\includegraphics[width=0.5\linewidth]{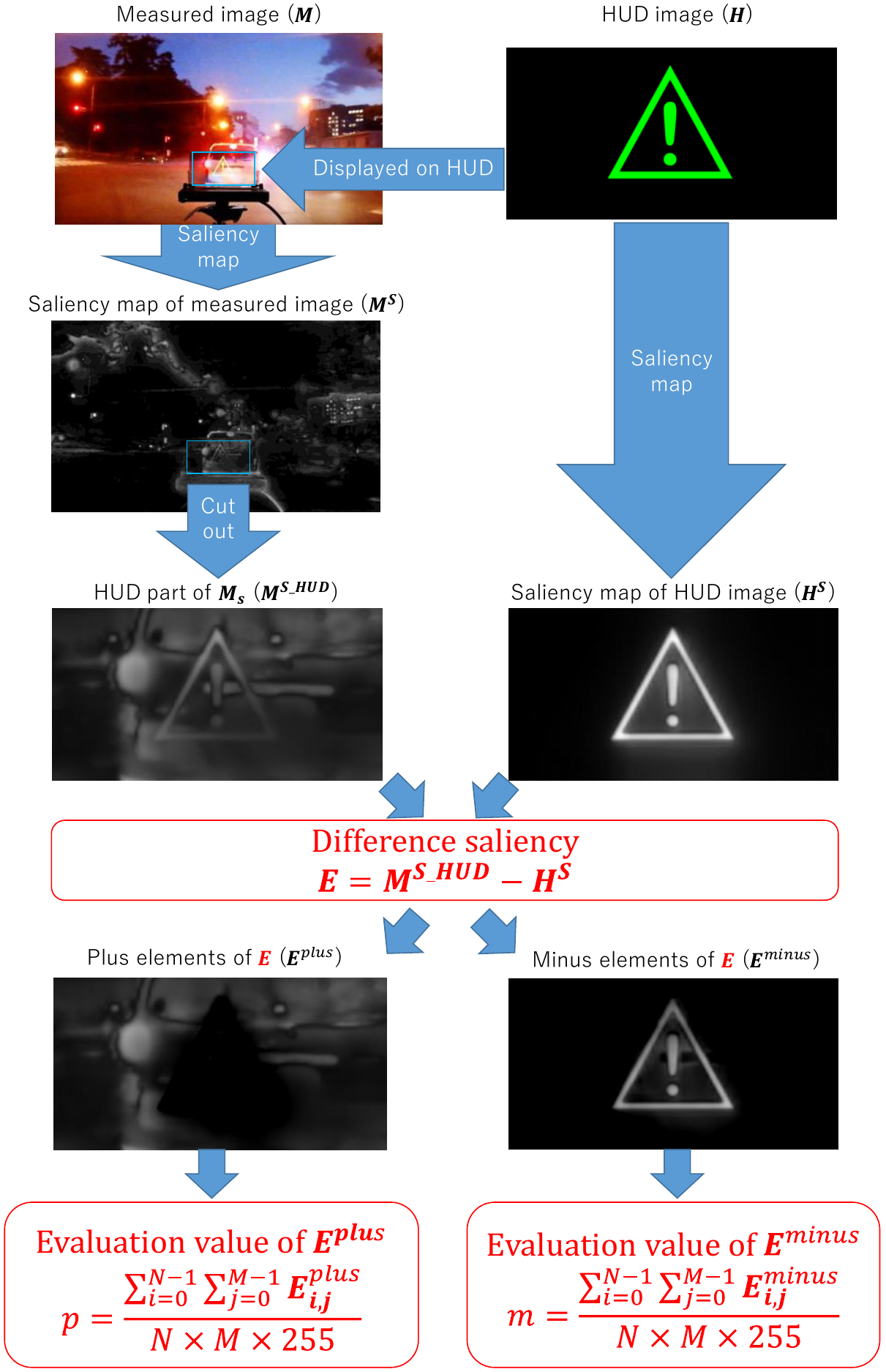}
\vspace{-2mm}
  \caption{Difference saliency method}
  \label{method}
\end{figure}

This paper proposes two indices to evaluate the mutual interference of saliency between the HUD and background.
The first index is a degree of reduction in the saliency of the projected information by the effect of the background on the HUD. 
The second one is a degree of visual distraction by the effect of the background on the HUD. 

The general process of the proposed method is shown in Fig.~\ref{method}.
First, the whole of the driver's view when using the HUD is called a measured image ${\bf M}\in\mathbb{Z^+}^{W\times H \times C}$, where $W$ and $H$ are the width and height of the measured image, and $C$ is the color channel of the measured image.
Then, a saliency map ${\bf M}^S\in\mathbb{Z^+}^{W\times H}$ is computed from ${\bf M}$.

After that, the HUD part is cut out from the ${\bf M}^S$ because this part is a target to be evaluated.
It is described as ${\bf M}^{S\_HUD}\in\mathbb{Z}^{N\times M}$, where
$N$ and $M$ are the height and width of ${\bf M}^{S\_HUD}$. 
The input image of the HUD is called a HUD image~${\bf H}\in\mathbb{Z}^{N\times M\times C}$.
In this paper, a background of the HUD image is set to be black because of the following two reasons.
First, the black color which is projected on the HUD will become transparent, and the transparency is one of the most advantages of the HUD. 
Second, this study considers that the saliency of colored information on the black background is relatively high. 

Next, the saliency map of the HUD image is generated and resized consistently with the size of ${\bf M}^{S\_HUD}$.
It is denoted as ${\bf H}^S\in\mathbb{Z}^{N\times M}$.
It will be used as a baseline to evaluate the mutual effect between the HUD and the background, and therefore the difference ${\bf E}$ between ${\bf M}^{S\_HUD}$ and ${\bf H}^S$ is calculated by
\begin{eqnarray}
  {\bf E} & = & {\bf M}^{S\_HUD}-{\bf H}^S\in\mathbb{Z}^{N \times M}.
\end{eqnarray}
The plus elements part of ${\bf E}$ is written as ${\bf E}^{plus}$, and the minus part is written as ${\bf E}^{minus}$.
Therefore, each element of those two parts is computed by
\begin{eqnarray}
  { E}^{plus}_{i,j} & = & \max({E}_{i,j}, 0 )
\\
  { E}^{minus}_{i,j} & = & -\min({ E}_{i,j},0)
\end{eqnarray}
Note that ${\bf E}^{plus}$ represents the saliency in the display scope of the HUD except for the projected information, and it 
is the visual distraction derived from the background.
The visual distraction will interfere with the driver in noticing the projected information on the HUD.
${\bf E}^{minus}$ represents a degree of reduction in the saliency of the projected information on the HUD. The large reduction in the saliency means that the background makes the driver feel difficult to notice the information on the HUD.

Final, ${\bf E}^{plus}$ and ${\bf E}^{minus}$ are converted to scalar values $p$ and $m$ as calculated by
\begin{eqnarray}
  p&=&\frac{\sum_{i=0}^{N-1}\sum_{j=0}^{M-1}{E}^{plus}_{i,j}}{N\times M\times 255},
\\
  m&=&\frac{\sum_{i=0}^{N-1}\sum_{j=0}^{M-1}{E}^{minus}_{i,j}}{N\times M\times 255},
\end{eqnarray}
which are defined as the two evaluation indices. 
Note that the saliency in the areas of the projected information on the HUD are different when the projected information are different.
Therefore, the indices can only evaluate the same information contents with different colors in the same background.

\section{Experiment}
A simulation experiment was divided into two parts.
The first part showed problems of the HUD used in different backgrounds, and the second part applied the proposed method to evaluate objectively the mutual interference of saliency between the HUD and background.

\subsection{Recording of driving video}
First, a video of the front scene was taken by a camera~(GoPro HERO+\footnote{GoPro HERO+: https://gopro.com/update/heroplus}) while a driver was driving a vehicle in an urban area of Nagoya, Japan.
The five-hour video includes scenes of different time periods such as daytime, evening, and night, and scenes of different traffic environments such as turning at an intersection, driving in a traffic jam, and driving on a crowded road with the large pedestrian flow. 
The video also recorded various glare conditions such as brake lamps of the preceding vehicles, headlights of the oncoming vehicle and sunlight.

\subsection{Experiment setting}
\begin{figure}[bt]
\centering
\begin{minipage}[t]{0.49\textwidth}
\centering
\includegraphics[width=1\linewidth]{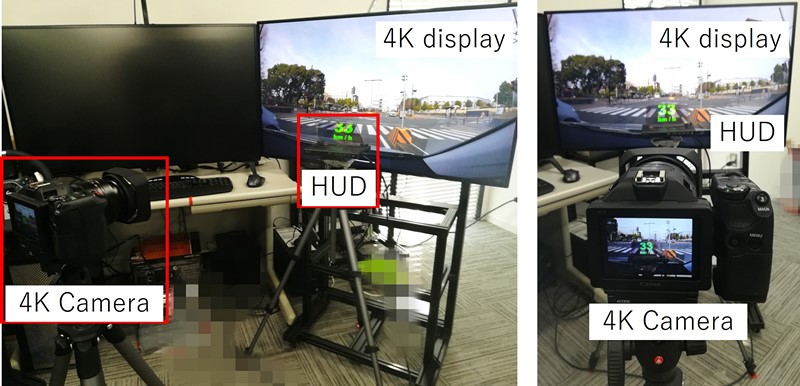}
\vspace{-2mm}
  \caption{Experimental equipment.}
  \label{experiment}
\end{minipage}
\hspace{1mm}
\begin{minipage}[t]{0.49\textwidth}
\centering
\includegraphics[width=1\linewidth]{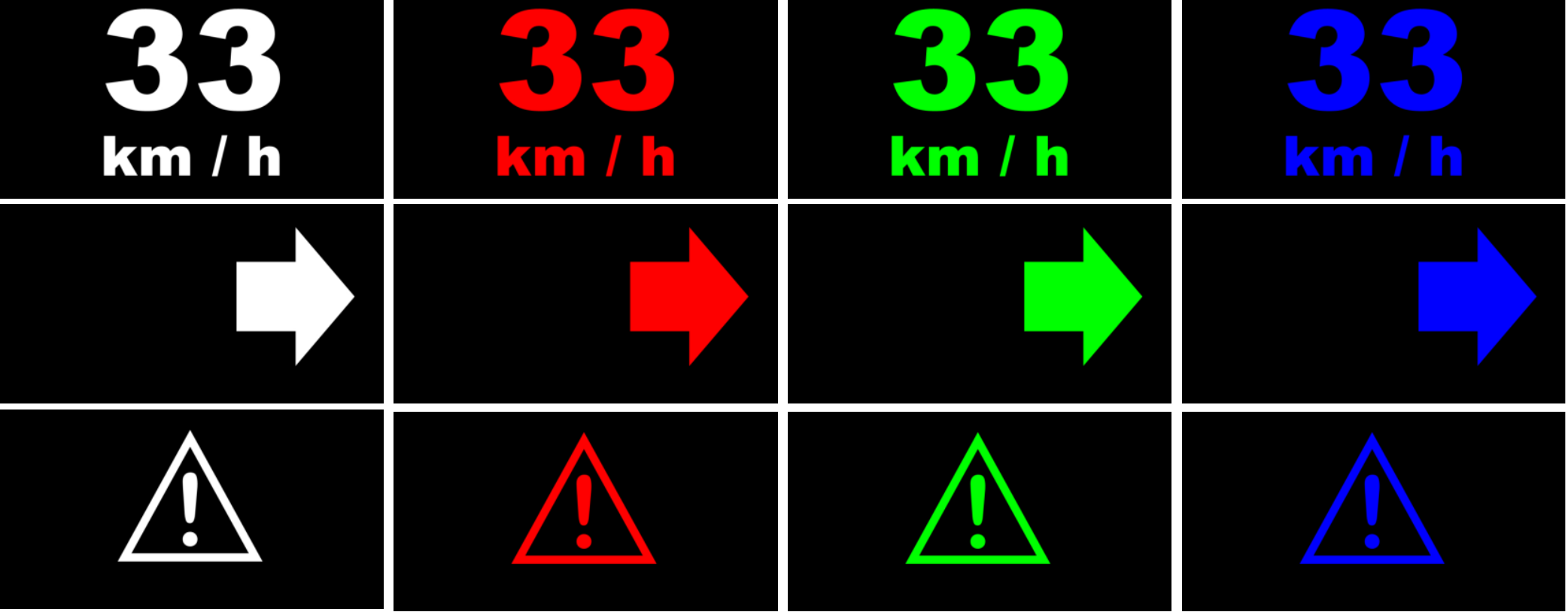}
\vspace{-2mm}
  \caption{The prepared information of the HUD.}
  \label{icon}
  \end{minipage}
\end{figure}

Figure~\ref{experiment} shows the experimental equipment.
The still images of the video was replayed on a 42.5-inch 4K display~(LG 43UD79T-B\footnote{LG 43UD79T-B: https://www.lg.com/jp/monitor/lg-43UD79T-B}) during the experiment conducted in a laboratory.
A 6.2-inch HUD~(MAXWIN HUD-622\footnote{MAXWIN HUD-622: https://www.maxwin.jp/content/hud/hud-622.html}) was placed in front of the 4K display to simulate the actual driving environment. 
A 4k camera~(Canon XC10\footnote{Canon XC10: https://cweb.canon.jp/prodv/lineup/xc10/}) was used to take photos in the case when overlap appears between information on the HUD and background.
The view angle, elevation angle, and the height of the camera are set close to the real driving vision of the driver.

There were three kinds of information contents that were projected on the HUD, as shown in Fig.~\ref{icon}.
Each of the information has four colors: white, red, green, and blue.

In this experiment, the most basic saliency method proposed by \cite{itti1998model}~\footnote{A open source of Itti's saliency method was used in the experiment: https://github.com/akisato-/pySaliencyMap} was used.
Note that the motion-based saliency was not considered and the RGB color space was used in this experiment.

\subsection{Problems of the HUD used in different backgrounds}

\begin{figure}[tb]
\centering
\begin{minipage}[t]{0.49\textwidth}
\centering
\includegraphics[width=1\linewidth]{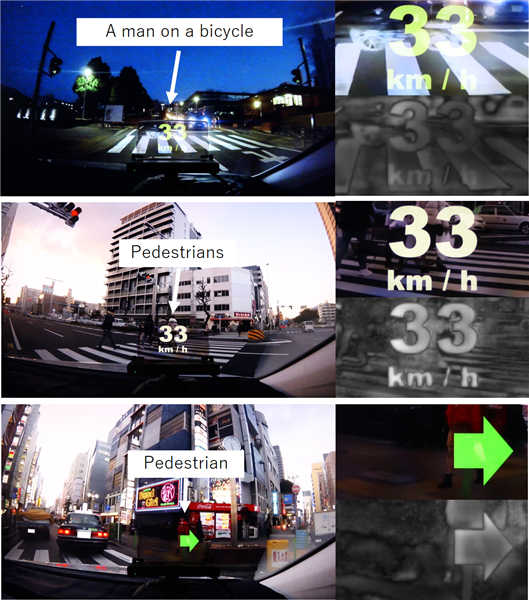}
\vspace{-2mm}
  \caption{The projected information on the HUD interfered the driver to observe objects in the environment.}
  \label{e01}
\end{minipage}
\hspace{1mm}
\begin{minipage}[t]{0.49\textwidth}
\centering
\includegraphics[width=1\linewidth]{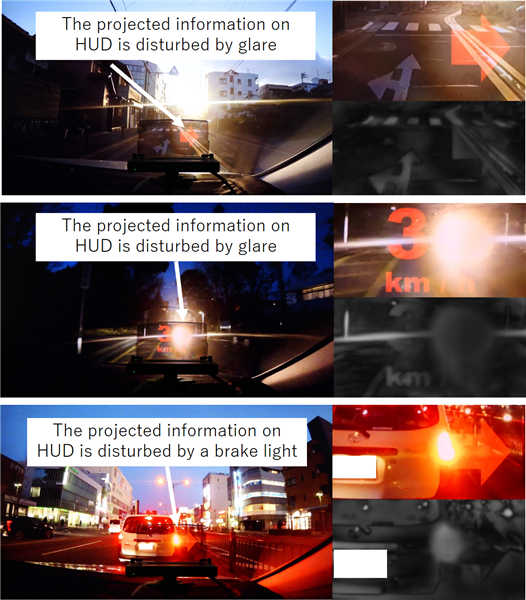}
\vspace{-2mm}
  \caption{The ambient light attenuated the saliency of the projected information on the HUD.}
  \label{e02}
    \end{minipage}
\end{figure}

There were two typical scenes to show the mutual interference between the information on the HUD and the background.
The first scene is that the projected information on the HUD is too eye-catching, thereby it will interfere with the driver in observing objects from the surrounding environment. 
Three examples are shown in Fig.~\ref{e01}.
In each example, the left image shows the whole of the scene from the driver's viewpoint; the upper-right image is the HUD part cut from the left image; the bottom right image is the HUD part cut from the saliency map of the left image.
These three examples showed that the pedestrians were blocked either during the night-time or the daytime.
The HUD part cut from the saliency map shows that the saliency of the projected information on the HUD was higher than the saliency of the pedestrian part.
This situation has a high possibility to increase the risk of an accident when the driver is attracted by the information on the HUD.

The second scene is that the saliency of the projected information on the HUD will be reduced in the case of glare.
The glare refers to the light not only from the sun but also from the brake lights of the preceding vehicles and the headlights of the oncoming vehicles.
Three typical examples are shown in Fig.~\ref{e02}.
The HUD parts cut from the saliency map clearly show that the saliency of the projected information on the HUD was significantly lower than the saliency of the background.

\subsection{Objective evaluation experiment}
This subsection evaluated objectively the saliency of different color information on the HUD in the same background. 
There are two scenes shown as examples in Table~\ref{table:result}.
One of the scenes is in the daytime and the other is at night-time.
For the HUD images, the same information with different colors had a very similar saliency map, although the saliency of blue on a black background is lower than others by human vision.
There are two reasons for this:
1) \cite{itti1998model}'s model normalizes feature maps from each color channels.
2) The distances from red (255, 0, 0), green (0, 255, 0), blue (0, 0, 255) to the black background color (0, 0, 0) in RGB space are the same.

The objective evaluation results by the proposed method are shown in Table~\ref{table:result} as the plus value $p$ and the minus value $m$.
The minus values $m$ of the red information were higher than other colors in the both of two scenes. It means that the saliencies of the red information were reduced because the red brake light of the preceding vehicle and the red information on the HUD overlapped each other.
The plus value $p$ shows the visual distraction of information on the HUD.
In the each scene, plus values were similar even the colors of information were different.
Comparing with the ${\bf E}^{plus}$ of two scenes in the Table~\ref{table:result}, 
the white areas show the visual distraction from the background.
The visual distraction of the upper scene was strong than it of the lower scene because the the white area of the upper scene is larger and more pronounced than the the lower scene.
Thus, the plus values $p$ of the upper scene were higher than the plus values $p$ of the lower scene.
Above results were consistent with the assumptions in this paper.

\section{Conclusion}
In this paper, a difference saliency method was proposed for objectively evaluating the mutual effect of HUD with various background.
It can be used for any saliency map method.
A simulation experiment was performed to clarify that the saliency of information on the HUD was reduced by the light conditions (glare from the sun, the brake light of the front vehicle, the headlights in the opposite lane).

The red information should be carefully used in the HUD contents because the saliency of a red color often becomes weaker than other colors in some scenes.
The reason for this is that there are many warm light sources in the driving environment, such as brake lights, sunsets, and yellowed street lights at night.
Consequently, the results of this experiment indicate that it is better not to use the red color to provide emergency information on the HUD in the vehicle although the red color is usually used to show the dangerous information in the human-machine interface.

In the future, the proposed method will be extended to evaluate the saliency of a movie by extracting the motion feature. Moreover, other color spaces that are closer to human vision, such as LUV and LAB, will be applied in the proposed method.
\bibliographystyle{IEEEtran}
\bibliography{main}

\begin{landscape}
\begin{center}
\begin{table}[p]
 \caption{Objective saliency evaluation for the same information projected on the HUD with different colors in the same background.}
 \label{table:result}
\begin{center}
  \begin{tabular}{ccccccccc}
 \hline
   \tabincell{c}{Measured image\\(${\bf M}$)}&
    \tabincell{c}{Saliency map of\\measured image\\(${\bf M}^S$)}&
    \tabincell{c}{HUD part of ${\bf M}_s$\\(${\bf M}^{S\_HUD}$)}&
    \tabincell{c}{HUD image\\(${\bf H}$)}&
    \tabincell{c}{Saliency map of\\HUD image\\(${\bf H}^s$)}&
    \tabincell{c}{Plus elements of\\${\bf M}^{S\_HUD}-{\bf H}^s$\\(${\bf E}^{plus}$)}&
    \tabincell{c}{Minus elements of\\${\bf M}^{S\_HUD}-{\bf H}^s$\\(${\bf E}^{minus}$)}&
    \tabincell{c}{Plus\\value\\(p)}& 
    \tabincell{c}{Minus\\value\\(m)}\\
     \hline
     \multicolumn{9}{c}{Case~1: an example of using HUD during the daytime}\\\hline
   \scalebox{1}{\includegraphics[width=27mm]{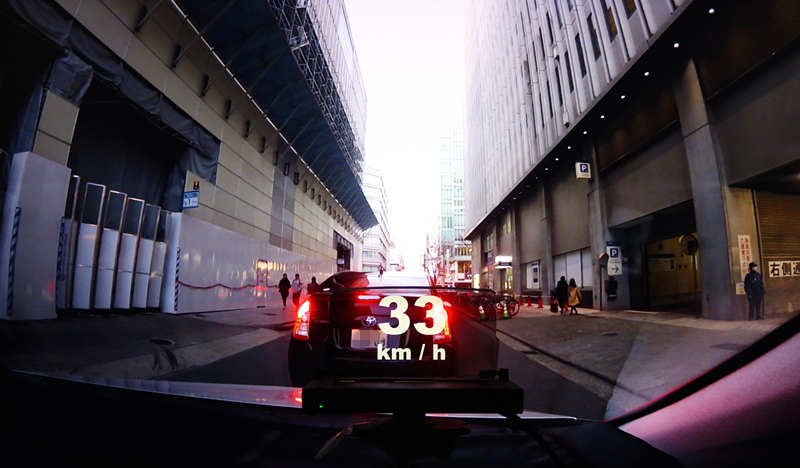}}&
     \scalebox{1}{\includegraphics[width=27mm]{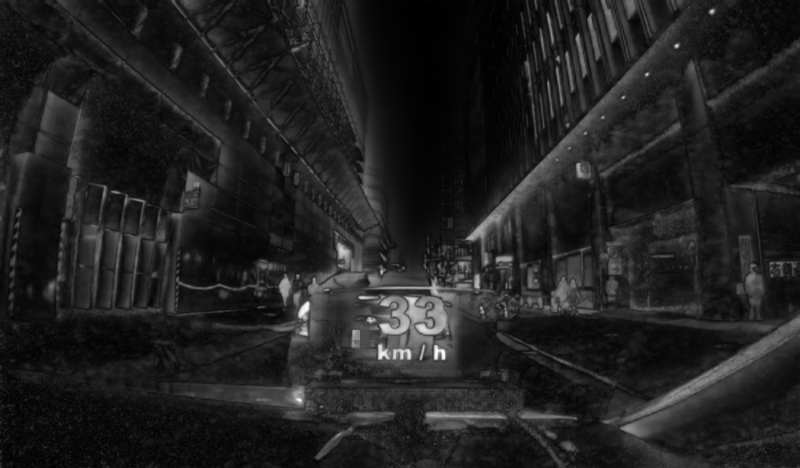}}& 
     \scalebox{1}{\includegraphics[width=27mm]{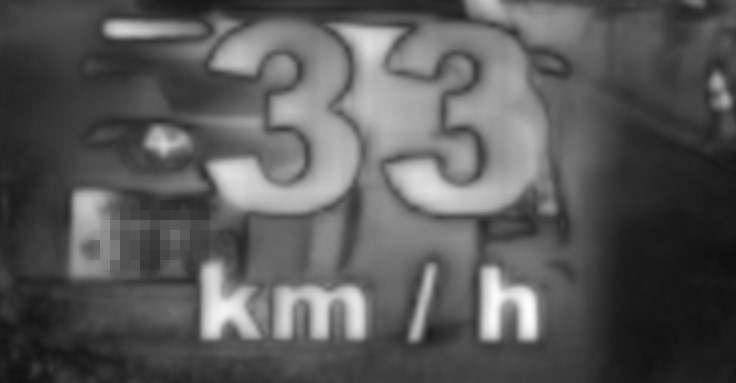}}&
     \scalebox{1}{\includegraphics[width=27mm]{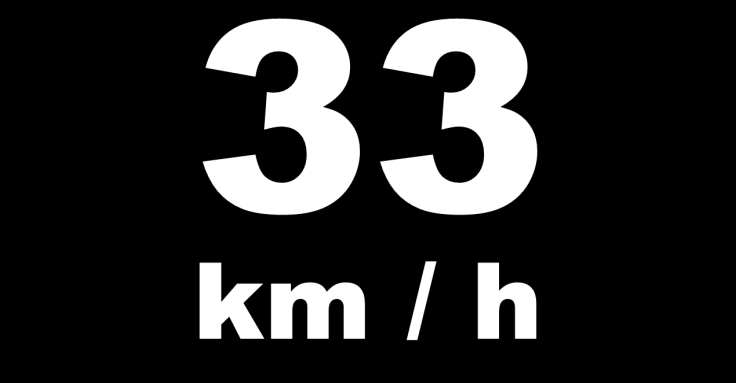}}&
     \scalebox{1}{\includegraphics[width=27mm]{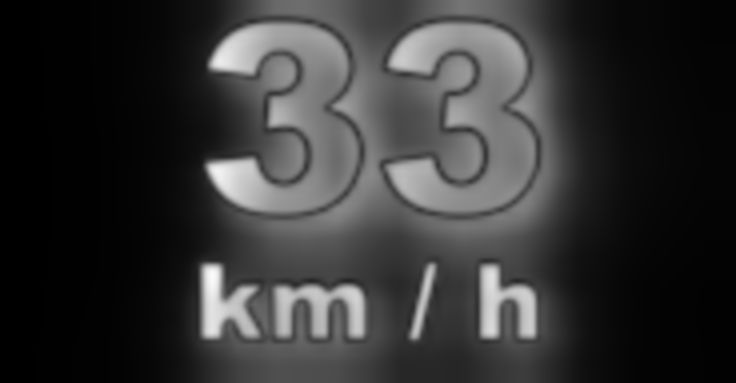}}&
     \scalebox{1}{\includegraphics[width=27mm]{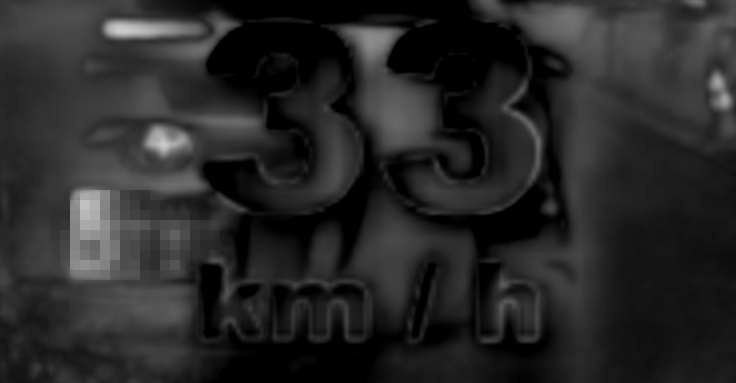}}&
     \scalebox{1}{\includegraphics[width=27mm]{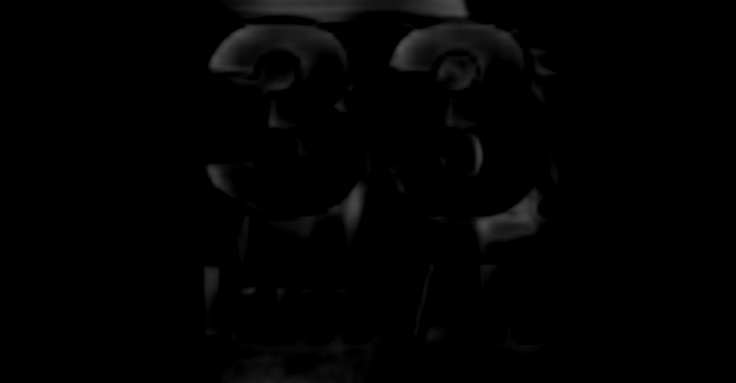}}&
     0.156 &0.009\\ \hline
     
        \scalebox{1}{\includegraphics[width=27mm]{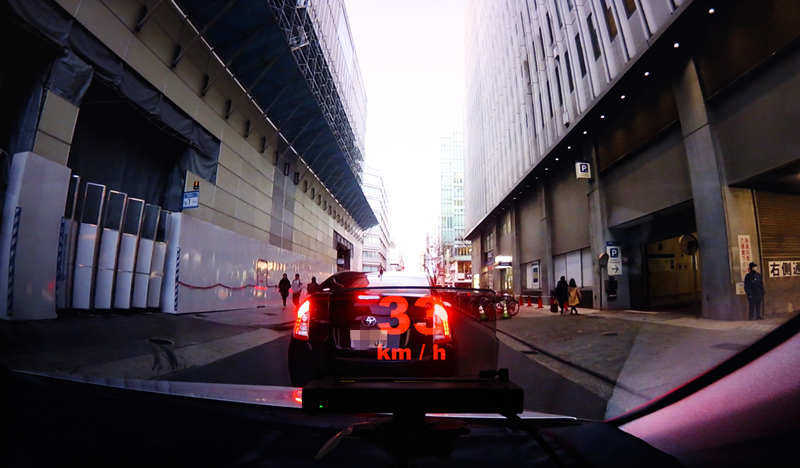}}&
     \scalebox{1}{\includegraphics[width=27mm]{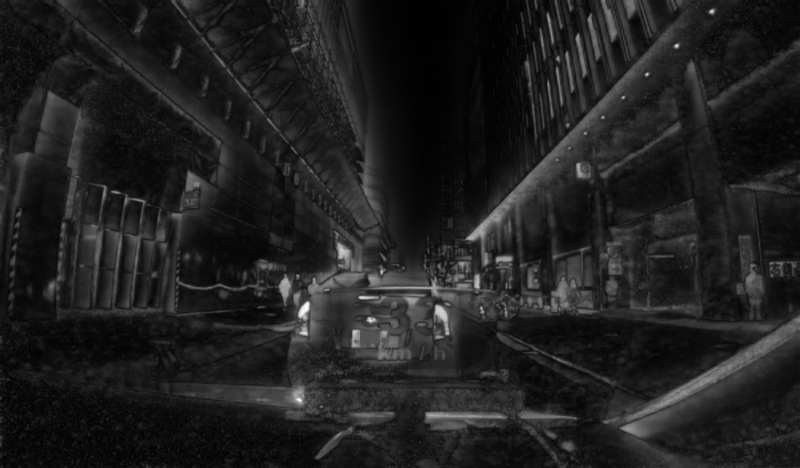}}& 
     \scalebox{1}{\includegraphics[width=27mm]{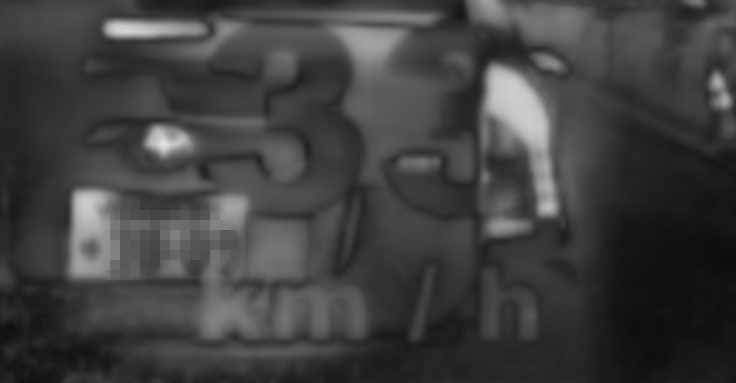}}&
     \scalebox{1}{\includegraphics[width=27mm]{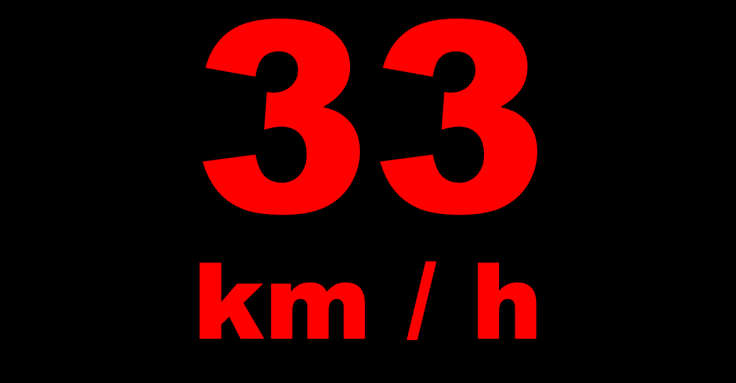}}&
     \scalebox{1}{\includegraphics[width=27mm]{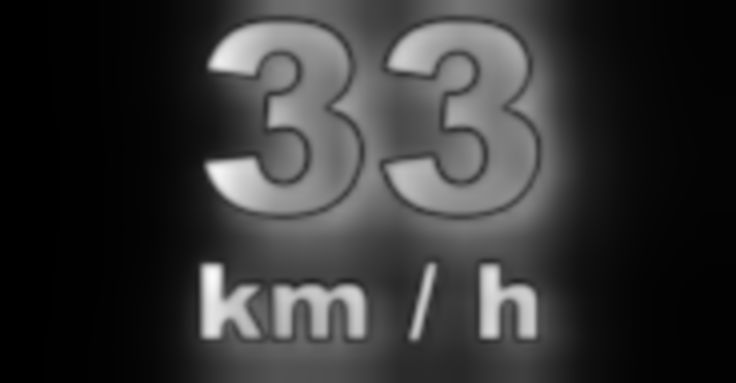}}&
     \scalebox{1}{\includegraphics[width=27mm]{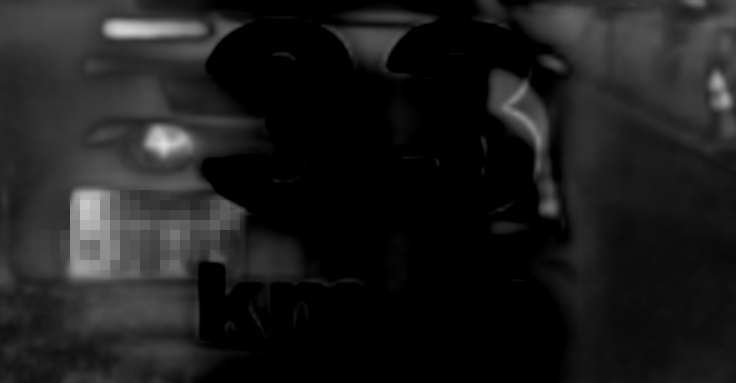}}&
     \scalebox{1}{\includegraphics[width=27mm]{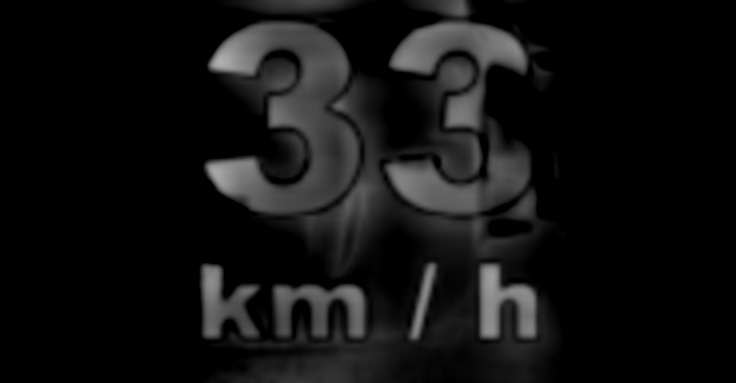}}&
     0.113 & 0.067 \\ \hline
     
        \scalebox{1}{\includegraphics[width=27mm]{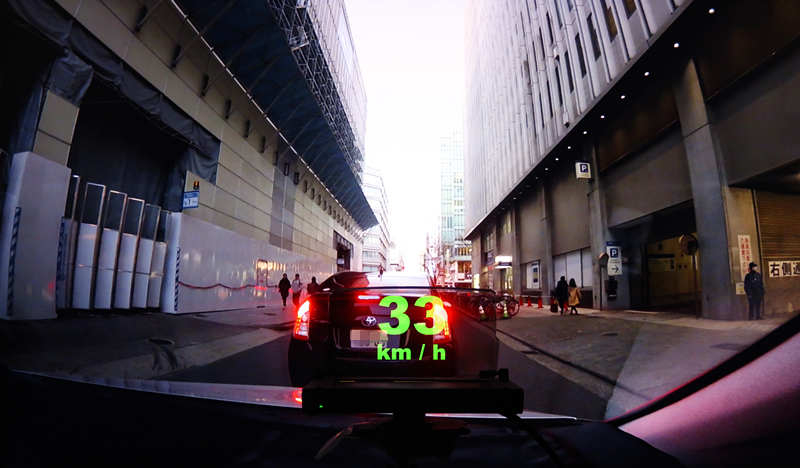}}&
     \scalebox{1}{\includegraphics[width=27mm]{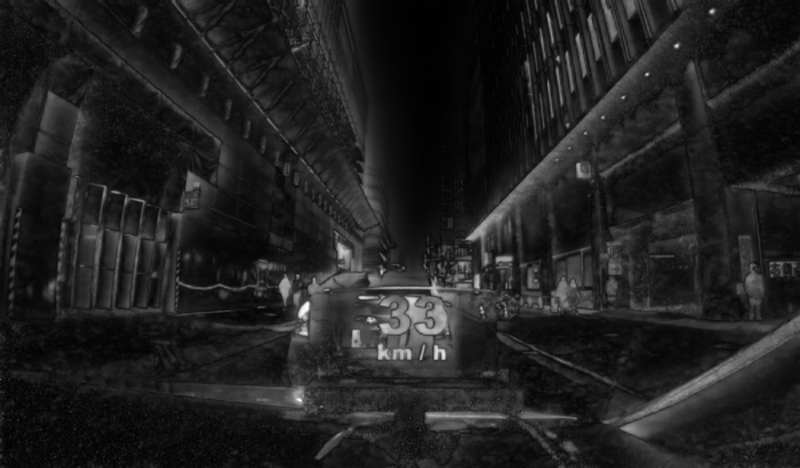}}& 
     \scalebox{1}{\includegraphics[width=27mm]{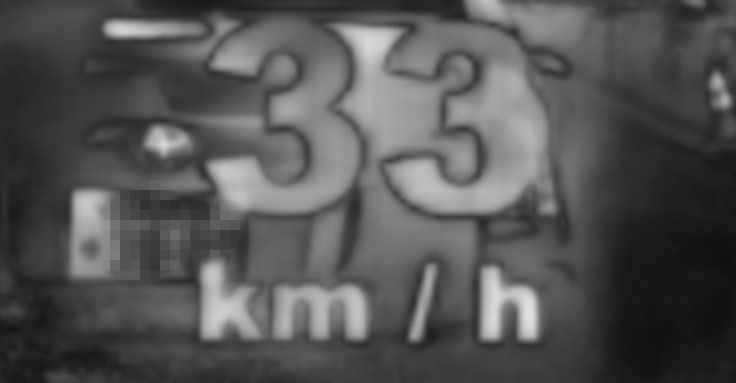}}&
     \scalebox{1}{\includegraphics[width=27mm]{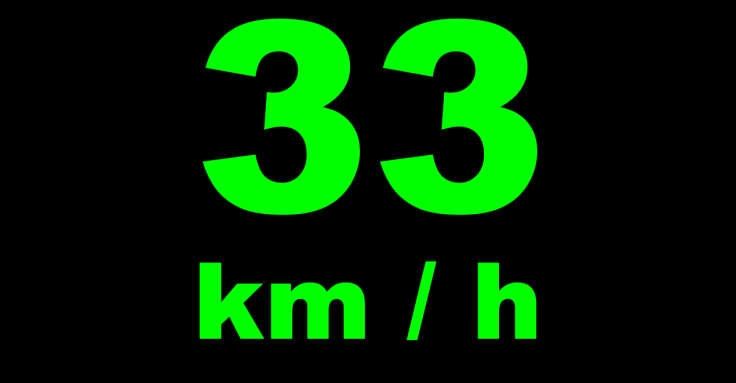}}&
     \scalebox{1}{\includegraphics[width=27mm]{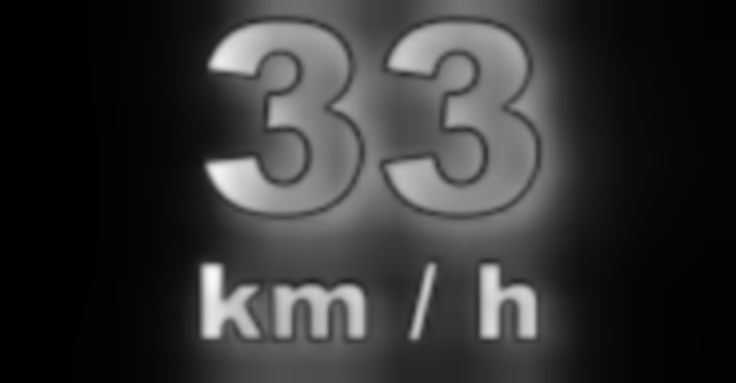}}&
     \scalebox{1}{\includegraphics[width=27mm]{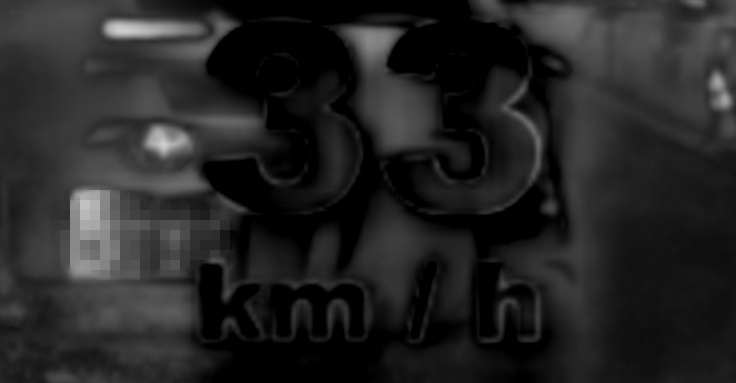}}&
     \scalebox{1}{\includegraphics[width=27mm]{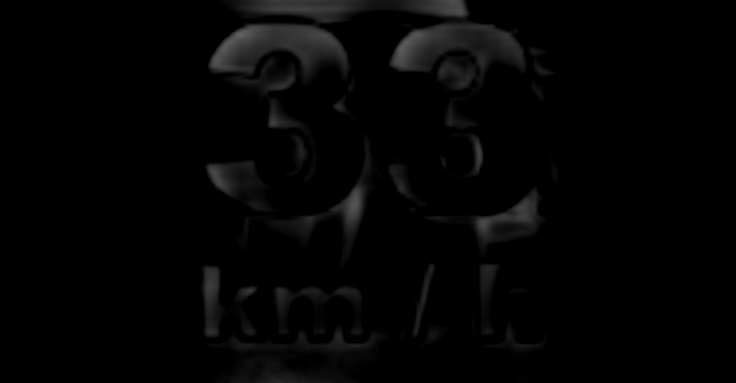}}&
     0.143  & 0.016 \\ \hline

     \scalebox{1}{\includegraphics[width=27mm]{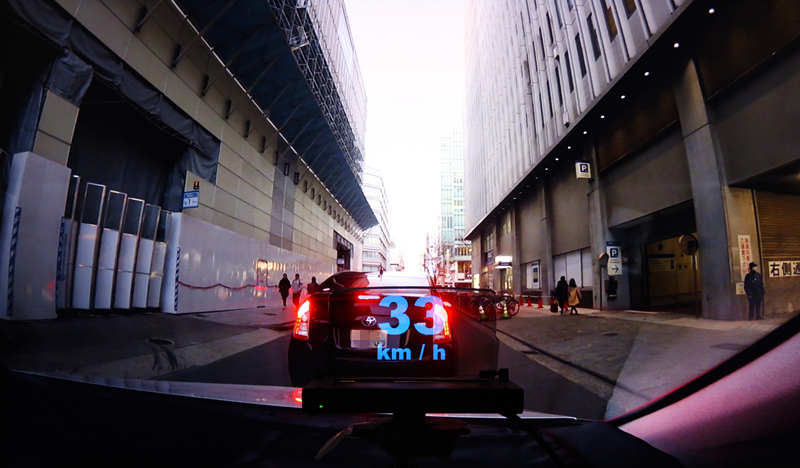}}&
     \scalebox{1}{\includegraphics[width=27mm]{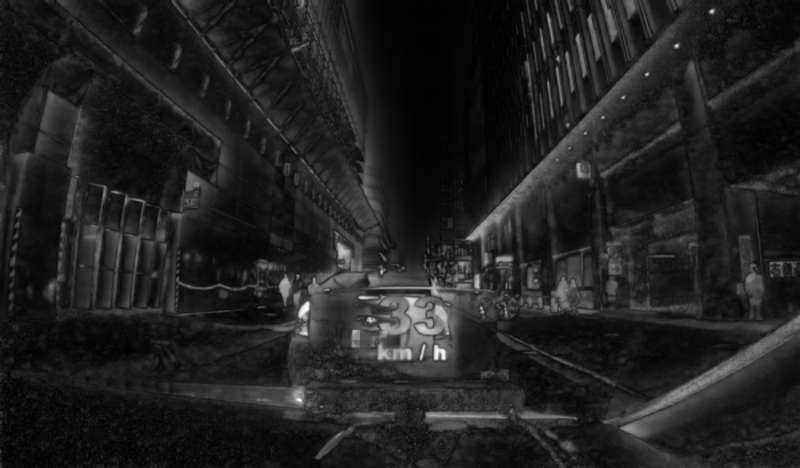}}& 
     \scalebox{1}{\includegraphics[width=27mm]{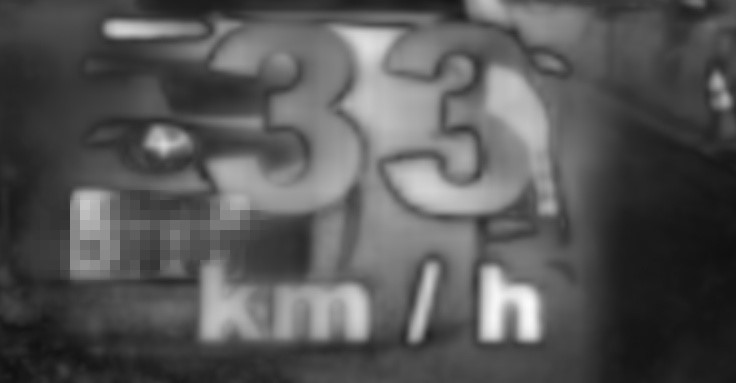}}&
     \scalebox{1}{\includegraphics[width=27mm]{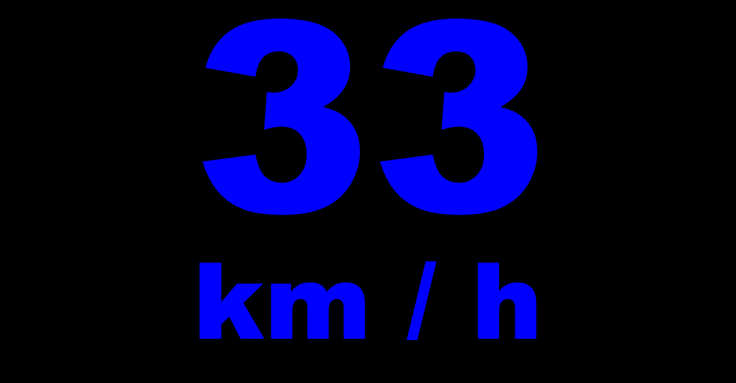}}&
     \scalebox{1}{\includegraphics[width=27mm]{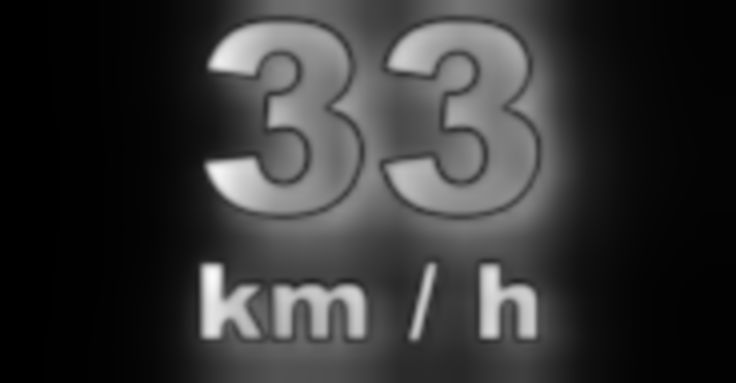}}&
     \scalebox{1}{\includegraphics[width=27mm]{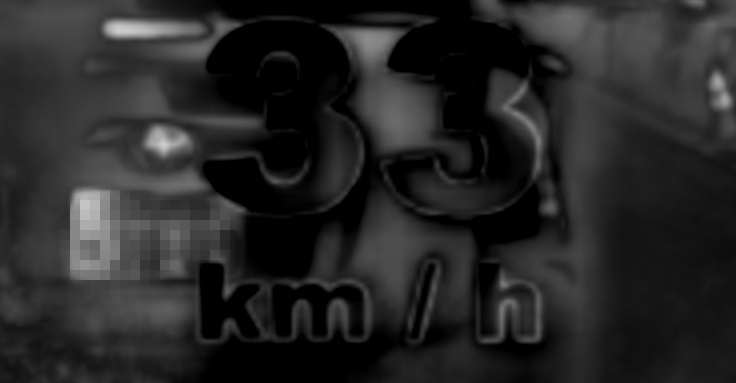}}&
     \scalebox{1}{\includegraphics[width=27mm]{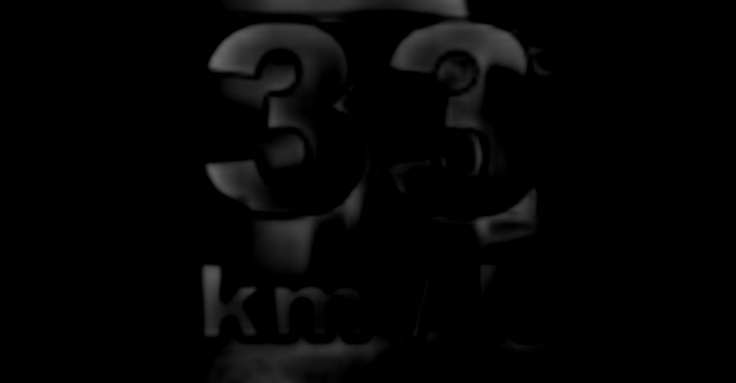}}&
     0.153 & 0.019 \\ \hline 
      \multicolumn{9}{c}{Case~2: an example of using HUD during the night}\\\hline
     
   \scalebox{1}{\includegraphics[width=27mm]{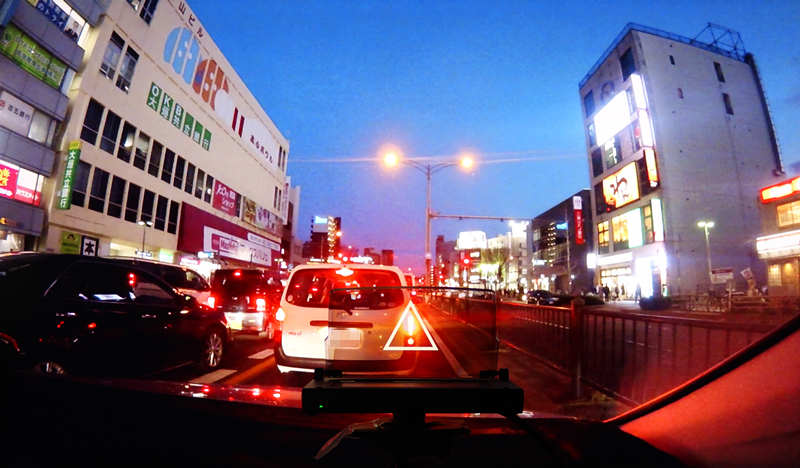}}&
     \scalebox{1}{\includegraphics[width=27mm]{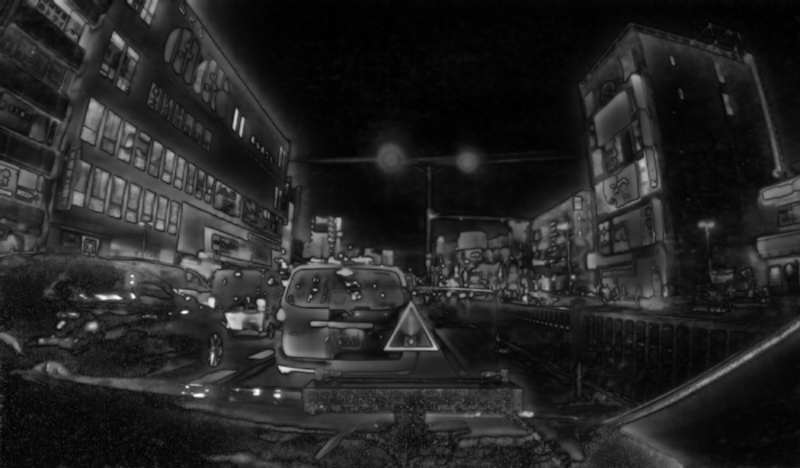}}& 
     \scalebox{1}{\includegraphics[width=27mm]{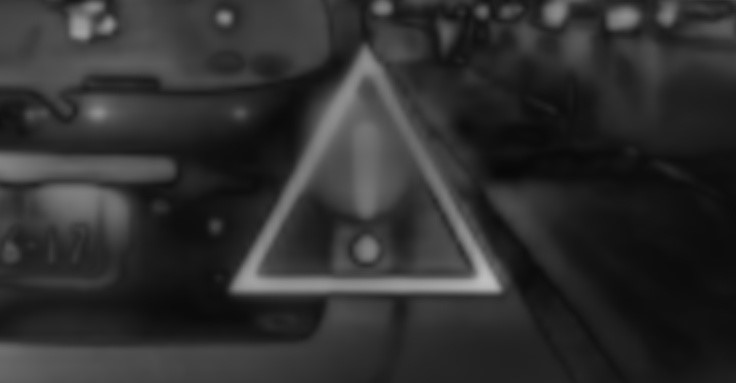}}&
     \scalebox{1}{\includegraphics[width=27mm]{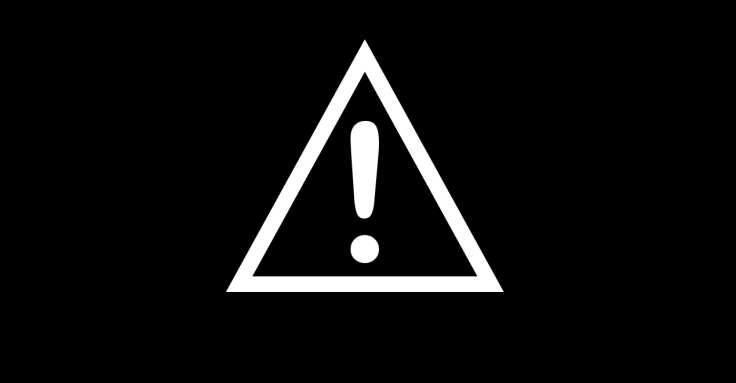}}&
     \scalebox{1}{\includegraphics[width=27mm]{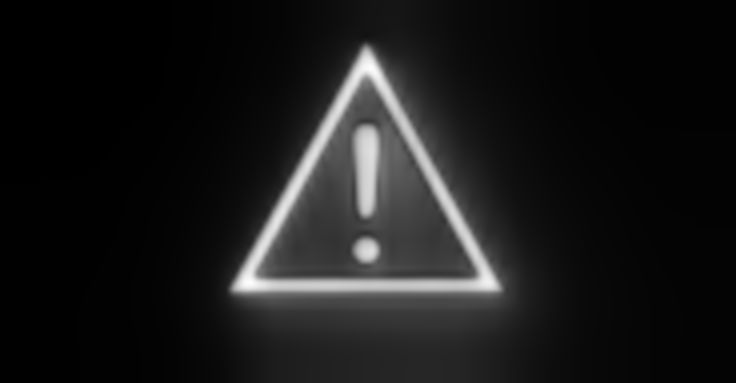}}&
     \scalebox{1}{\includegraphics[width=27mm]{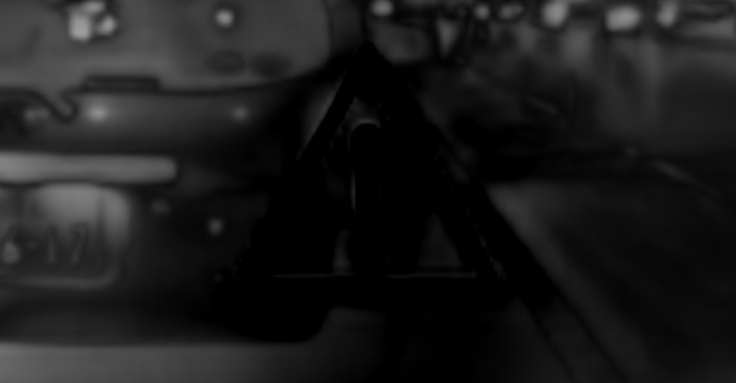}}&
     \scalebox{1}{\includegraphics[width=27mm]{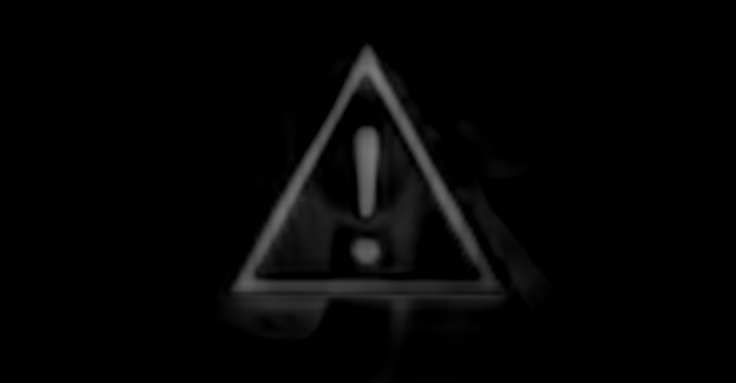}}&
     0.108 &0.020\\ \hline
     
        \scalebox{1}{\includegraphics[width=27mm]{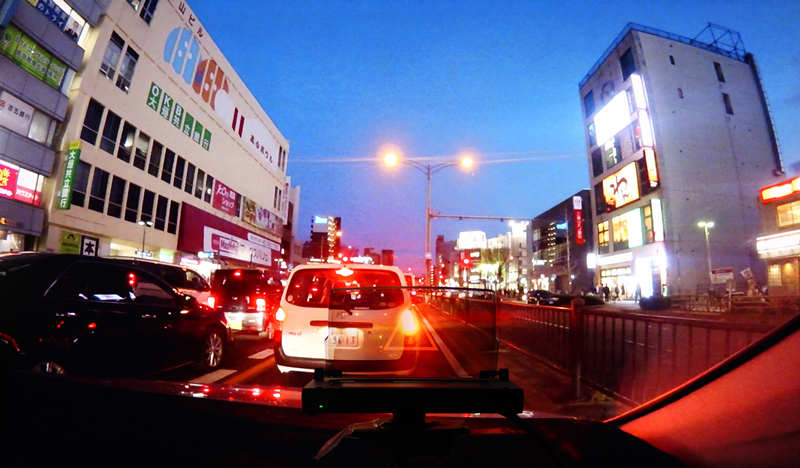}}&
     \scalebox{1}{\includegraphics[width=27mm]{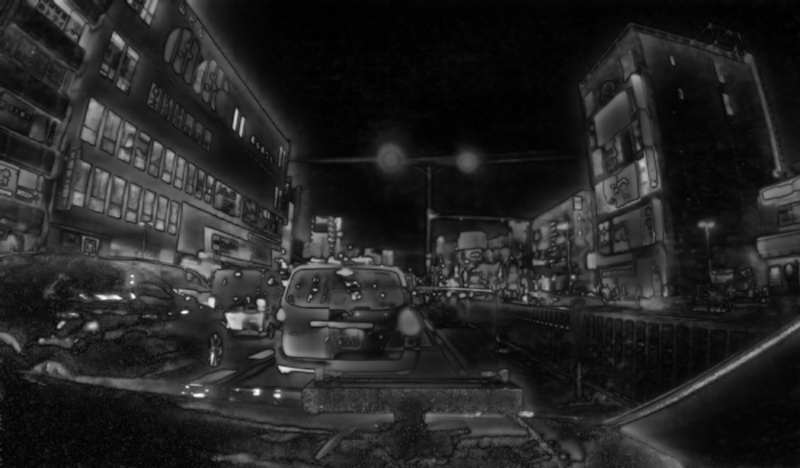}}& 
     \scalebox{1}{\includegraphics[width=27mm]{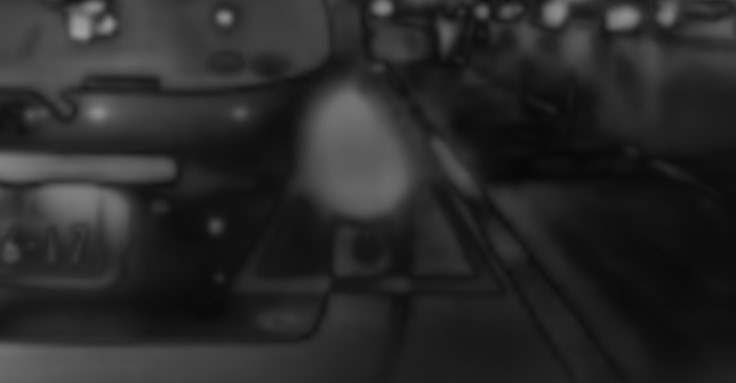}}&
     \scalebox{1}{\includegraphics[width=27mm]{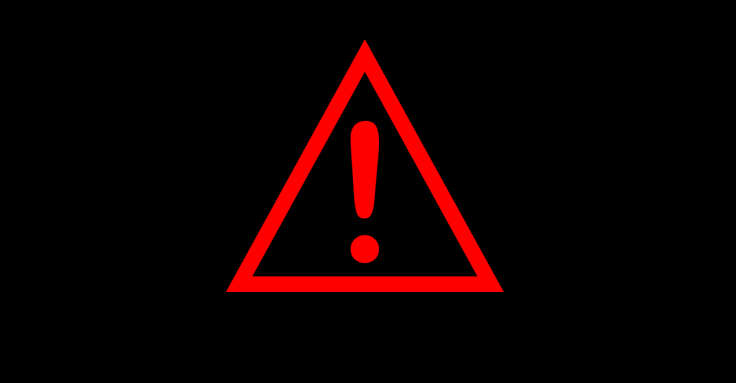}}&
     \scalebox{1}{\includegraphics[width=27mm]{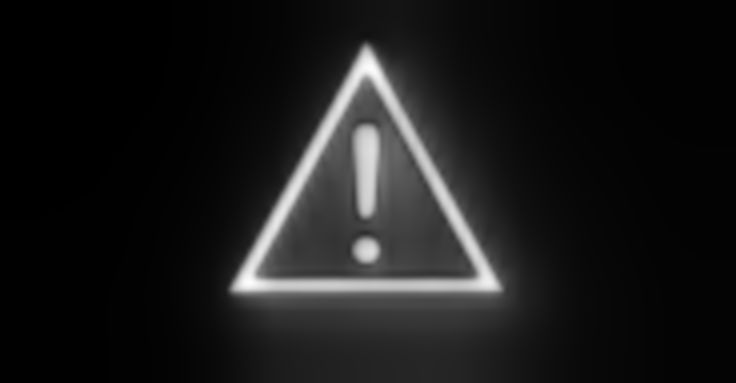}}&
     \scalebox{1}{\includegraphics[width=27mm]{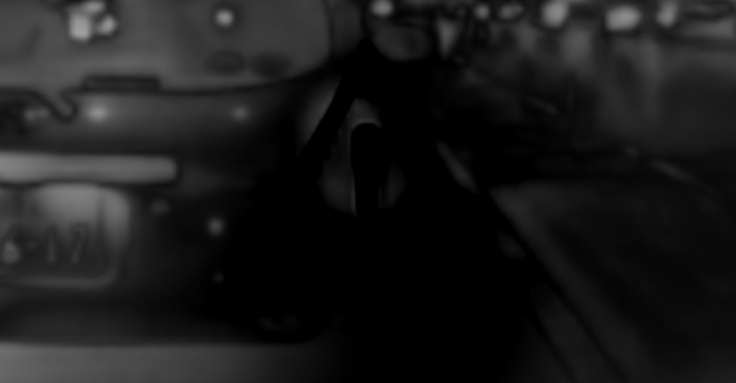}}&
     \scalebox{1}{\includegraphics[width=27mm]{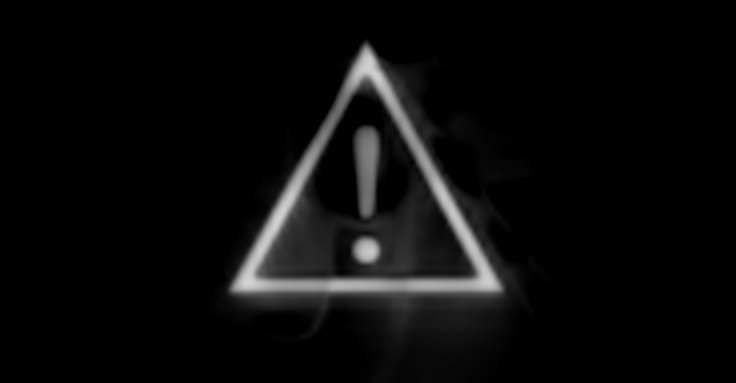}}&
     0.105 & 0.043 \\ \hline
     
        \scalebox{1}{\includegraphics[width=27mm]{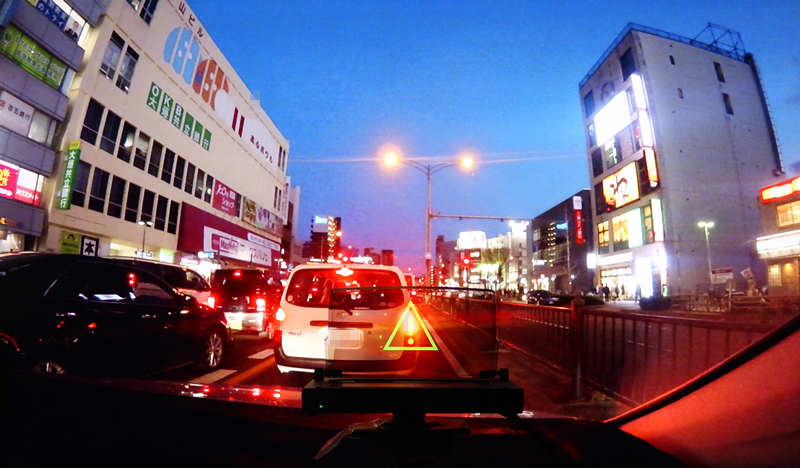}}&
     \scalebox{1}{\includegraphics[width=27mm]{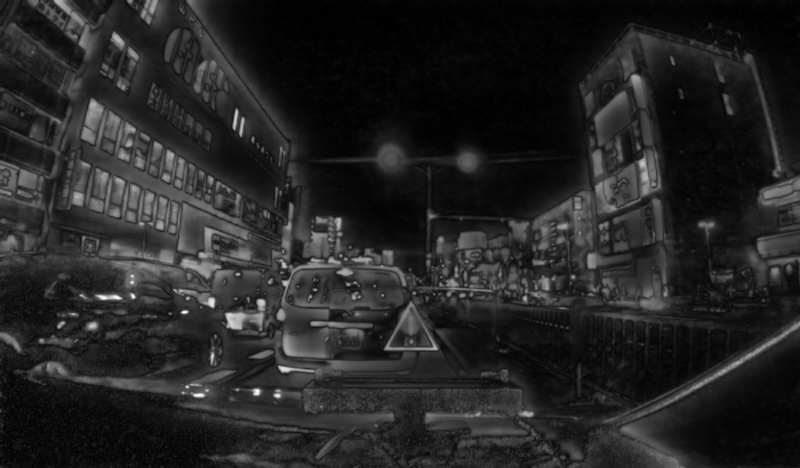}}& 
     \scalebox{1}{\includegraphics[width=27mm]{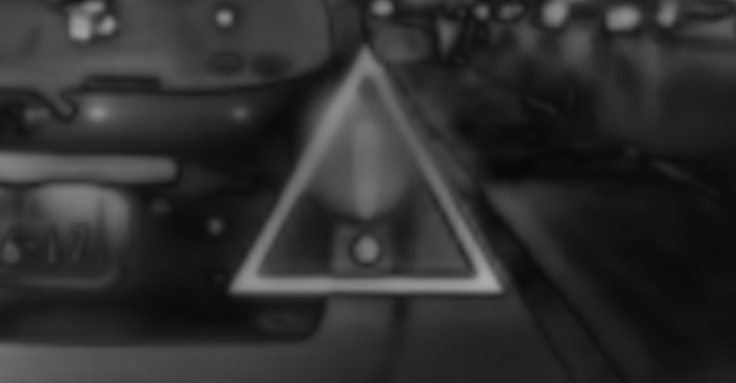}}&
     \scalebox{1}{\includegraphics[width=27mm]{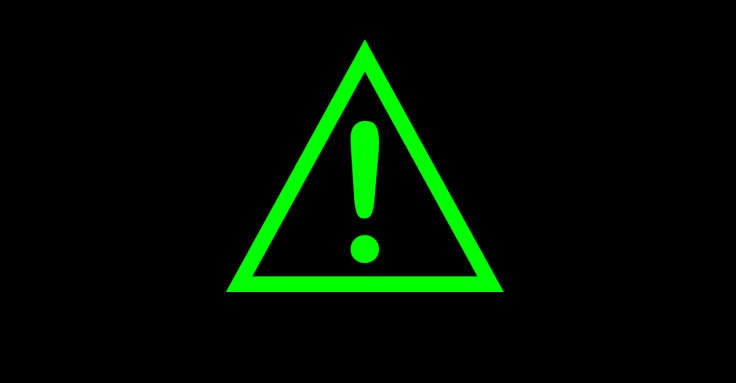}}&
     \scalebox{1}{\includegraphics[width=27mm]{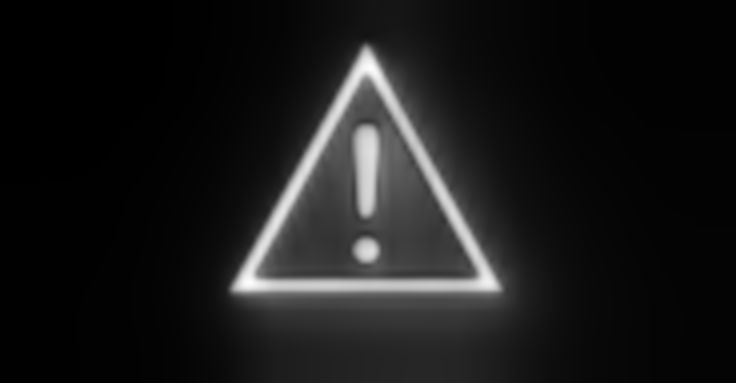}}&
     \scalebox{1}{\includegraphics[width=27mm]{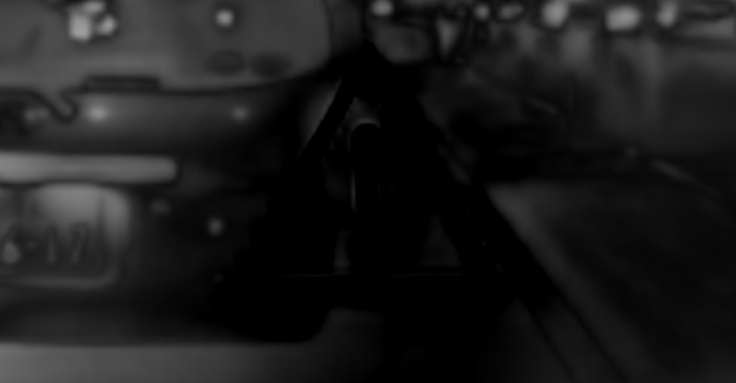}}&
     \scalebox{1}{\includegraphics[width=27mm]{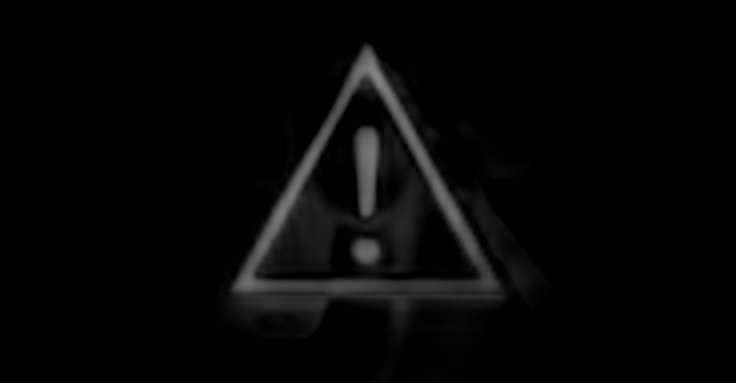}}&
     0.108  & 0.023 \\ \hline

     \scalebox{1}{\includegraphics[width=27mm]{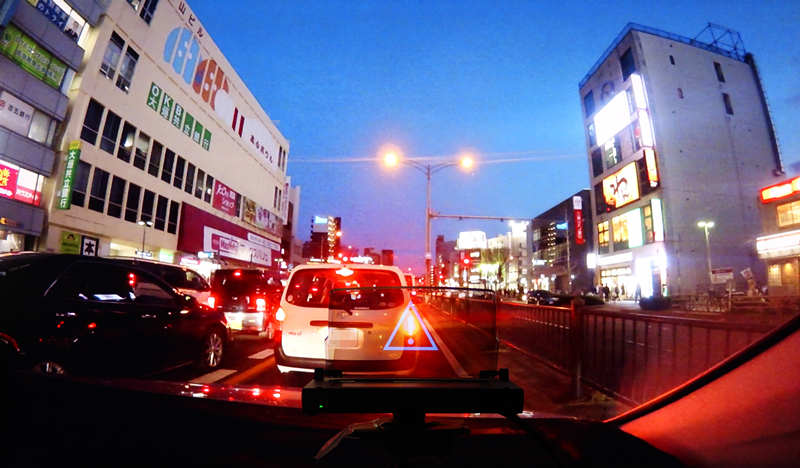}}&
     \scalebox{1}{\includegraphics[width=27mm]{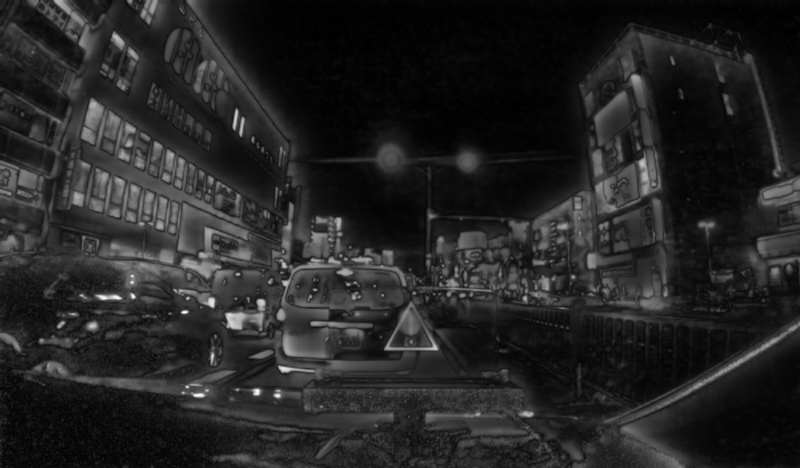}}& 
     \scalebox{1}{\includegraphics[width=27mm]{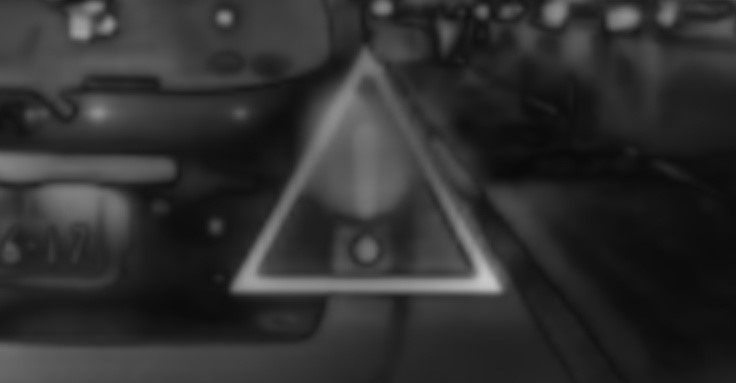}}&
     \scalebox{1}{\includegraphics[width=27mm]{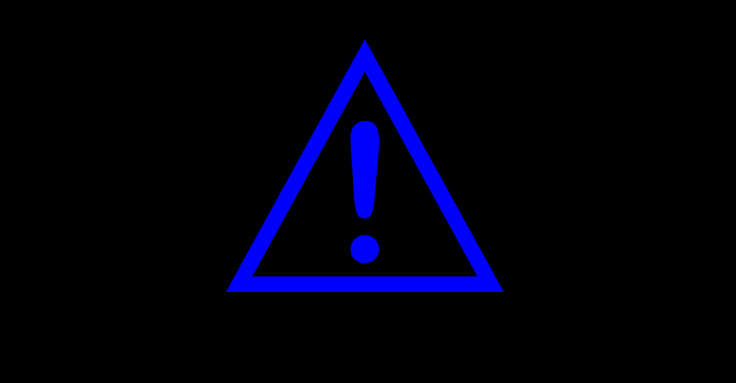}}&
     \scalebox{1}{\includegraphics[width=27mm]{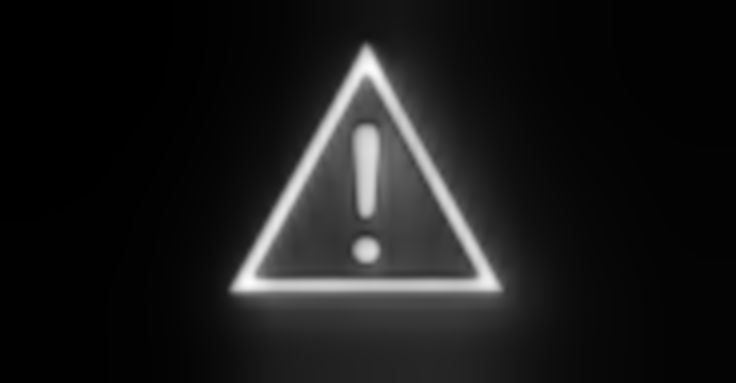}}&
     \scalebox{1}{\includegraphics[width=27mm]{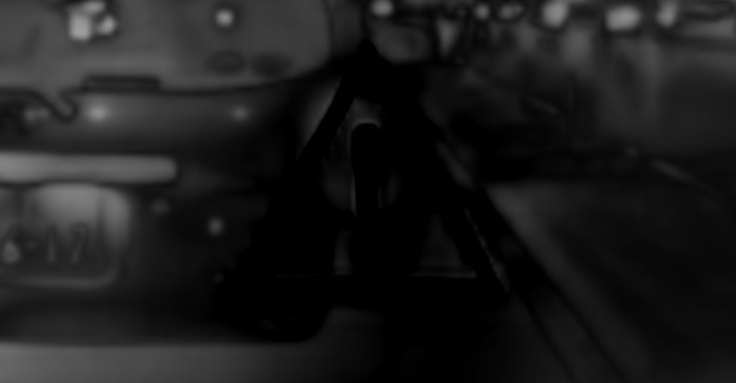}}&
     \scalebox{1}{\includegraphics[width=27mm]{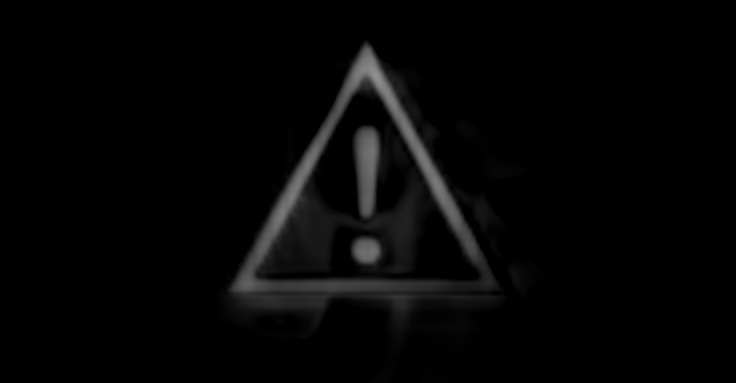}}&
     0.110 & 0.021 \\ \hline 
 \end{tabular}
\end{center}
\end{table}
\end{center}
\end{landscape}

\end{document}